# Reduced Symmetric 2D Photonic Crystal Cavity with Wavelength Tunability

Melike A. Gumus[1], Mediha Tutgun[1], Döne Yılmaz[1], and Hamza Kurt[1]

[1]Department of Electrical and Electronics Engineering, TOBB University of Economics and Technology, Ankara 06560, Turkey

**Abstract**

In this paper, we propose a microcavity supported by a designed photonic crystal (PhC) structure that supplies both tunability of cavity modes and cavity's quality factor. Low symmetric defect region provides a trigger effect for the frequency shifting by means of rotational manipulation of small symmetry elements. Deviation of effective filling ratio as a result of rotational modification within the defect region results in the emanation of cavity modes at different frequencies. Here, we numerically demonstrate the frequency shifting for each obtained mode with respect to defect region architecture. In addition to wavelength tunability, quality factor, mode volume, and Purcell factors are analyzed for the slightly modified structures. Also, electric field distributions of each mode that emerge at distinct frequencies have been also studied at adjusted frequency modes which are observed for all rotational modification scenarios as $\theta_{Rot} = [0°, 15°, 30°, 45°]$. After the investigations in 2D of silicon material ($\varepsilon_r = 12$), 3D simulations are performed and the collected data is used for the stacking approximation of 3D structures to get the 2D, thus the cross-checking of the quality factor that acquired from the 2D simulation can be executed by comparison with 3D. Limited 3D results are projected to approximate 2D ones step by step and get an exponential trend which reaches in the limit to the $\sim 10^8$ value for Q-factor. Besides, 2D and 3D simulations of alumina ($\varepsilon_r = 9.61$) in terms of mode analysis and quality factor have been repeated considering the microwave experiments. Therefore, experimental analysis is compared with the numerical results and good agreement between the two is found.

## 1. Introduction

Photonic crystals (PhC) are novel structures which have the ability of manipulation of light. A conceived PhC structure can manipulate the light behavior in many appealing ways and cavity effect is an intriguing one so that the surge of interest has increased in last decays. A PhC cavity provides a medium for the localization of light inside the structure by utilizing localized defects. Light is gathered into a small mode volume approximately in optical-wavelength dimensions whilst it is passing through the PhC, this behavior results of horizontal and vertical localization effect which is triggered by Distributed Bragg Reflection (DBR) and Total Internal Reflection (TIR). As is known, TIR can be described by the way of the *k* vector placement according to the light cone, which is directly associated with photonic band structure. In accordance with approximation, frequencies which take part above the light line are not allowed to propagate along the structure and they are known by the name of radiation modes. Whereas, the frequencies below the light line represents allowed modes and they named by guided modes. The contribution of the TIR to performance of any cavity is related to allowed mode portion. Vertically confined guided modes, resulting of TIR and they come to exist owing to defects belonging the structure, which are controllable for all structures. Otherwise, 3D structures can not exploit from the light due to failure to provide the TIR requirements totally, despite small mode volumes are still consistent, they sense this phenomenon as a loss mechanism and consequently these out of plane losses cause drastic decline of the vertical performance of the cavity according to 2D systems, which ensure the vertical localization [1–5]. However, the opportunities in the field of performance come at a price, fabrication irregularities and sustainability issues as; material and surface-state absorptions, surface roughness are the main obstacles for 2D PhCs [6]. These 2D microscale structures need rigorous fabrication processes to maintain the same performance and slight defect disarrangements cause serious losses. Several optical applications rely on TIR mostly, like microspheres or microdisks, besides in particularly 3D systems work by exposing to more configurable DBR mechanism [7]. Laterally confined guided modes are, emerged by the way of Bragg reflection of the light, from the surrounding PhC layers of the structure. Each layer of the PhCs employ like a mirror and a symmetric architecture provides an efficient localization effect from each direction of the cavity region. If the existing modes fulfill the requirements and they are confined laterally and vertically after disposing of defect region within a regular structure, generate a PhC cavity [8,9]. Applying structural modifications to the cavity region provides acquisition of targeted enhancement, as well as it is also a manner to control the cavity modes [10–13]. Besides regular and simple structures, complex ones have also investigated, like heterostructures or quasicrystal to acquire further cavity information [14–17]. Mentioned cavity studies above have opened up new opportunities to many of application for integrated devices such as filters [15,18], all-optical switches [19,20], sensors [21–24] and optical storing [25,26].

In the present work, two-dimensional square lattice PhC possessing defect region is investigated such that symmetry reduced configuration allows resonance mode tenability via rotational symmetrywithout any change in the filling ratio of the defect site. We will provide first numerical result both in frequency and time domains and microwave experimental verification is conducted.

## 2. Numerical Study of Low-symmetric photonic cavity

*2.1. Frequency Domain Analysis*

In this paper, we claim that the low symmetric PhC structures are good candidates to gather the cavity modes tunability owing to rotational tailoring without renounce of the performance of the cavity. In this stage, it has been observed that the equal *x* and *y* dimensions of structure help the light confinement and, enhance the cavity performance as a result of an effect which can be defined as symmetric mirror effect. By taking consider this clue, cavity design has been iterated on the way of structures that has equal x and y dimensions. Also, one of the further requirements to the occurrence of cavity effect is verification of an optimum layer number which surrounds the cavity region. After a series of the number of layer iterations, the optimum layer number is stated as $15a$ in this study. Eventually, the structural iterations which based on to find the ideal condition to establish a cavity region show that it has been achieved to the high-performance cavity design with a relatively compact array with $15a \times 15a$ size. The mentioned regular PhC structure with a low symmetric defect region is shown in figure 1(a), also the zoomed version is given in figure 1(b). As it is given, this region formed by four dielectric silicon rods with $R_1 = 0.20a$ and it is supported with smaller dielectric rods with $R_2 = 0.10a$ radii and $\varepsilon_r = 12$ dielectric constant, as a defect region. It is demonstrated that the active zone is designed on the basis of the smaller rods rotated getting the reference to the nearest bigger rods at specific angles, $\theta_{Rot} = [0°, 15°, 30°, 45°]$ in figure 1(c). Whilst the distance between the center of rotated small rods and the nearest bigger rod is $d = 0.35a$, it is $d = 2a$ between the center of the rods which are at the border of cavity region. Every rotation causes a new configuration and each of new configuration means that a different defect region is emerged within structure.

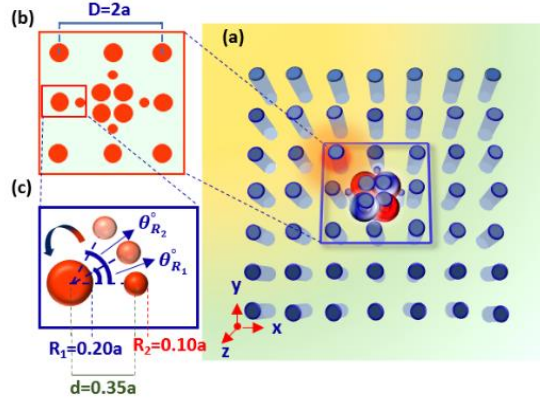

**Figure 1.** (a) Schematic diagram of 3D PhCs (a) cavity region which is consisted of defect region includes four big and four smaller silicon rods with $\varepsilon_r = 12$ and $R_1 = 0.20a$ and $R_2 = 0.10a$ radii, respectively. Also, the source has represented in the corner of the cavity region with red radiation. (b) Schematic representation of active zone including rotating rods based on nearest big rod with several specified angles are represented with $\theta_{Rot} = [0°, 15°, 30°, 45°]$ (c).

A defect region emanates some dispersion bands at the frequencies which support the cavity modes within the band gap of the structure. Mentioned dispersion bands are considered in the case of when they are below the light line, which is a limitation for k vectors according to the requirement of TIR $\boldsymbol{k_\parallel} > w/c$, where $\boldsymbol{k_\parallel}$ represent the in-plane wave vector, while $w$ and $c$ represents angular frequency and speed of light in vacuum, respectively. In the case of in-plane wave vector does not fulfill the TIR requirement, propagating light can not be capable to couple the structure effectively and this means that light leaks through the air as a loss. In keeping with this approximation 2D band diagram analyses which are performed by using MPB along $\Gamma - X$ symmetry points are represented in figure 2 for each of rotational manipulation of the defect region. The flat bands that are lying along the $\Gamma - X$ direction and below the light line clearly illustrate the approximately zero group velocity, $v_g = \partial w/\partial k \cong 0$, that means the light will localize within the cavity region over a satisfying time which results in a high quality factor [27]. Here, $v_g$ represents the group velocity whilst $w$ and $k$ defines angular frequency and wave vector, respectively. These flat bands show that there are four discrete cavity modes emerged within the bandgaps at

$a/\lambda_{0°} = [0.2876, 0.3598, 0.4061, 0.4125]$,

$a/\lambda_{15°} = [0.2876, 0.3566, 0.4103, 0.4116]$,

$a/\lambda_{30°} = [0.2877, 0.3523, 0.4163, 0.4089]$,

$a/\lambda_{45°} = [0.2877, 0.3511, 0.4128, 0.4082]$,

for each rotational case $\theta_{Rot} = [0°, 15°, 30°, 45°]$, respectively. The comments about the peak shifts of third and fourth modes have postponed after mode orders, which will be clear after transmission and mode profile investigations. It should be noted that there is a transition relation, which $a/\lambda = normalized\ frequency$, between the frequency domain to the time domain. Here, $a$ and $\lambda$ represents the lattice constant and the wavelength, respectively. In our numerical calculations, lattice constant value is $a = 561.6\ nm$ of the $15a \times 15a = 8.40\ \mu m \times 8.40\ \mu m$ sized relatively compact PhC structure. According to this relation, a convergence from the normalized frequency of bands to the wavelengths of the transmission can be possible.

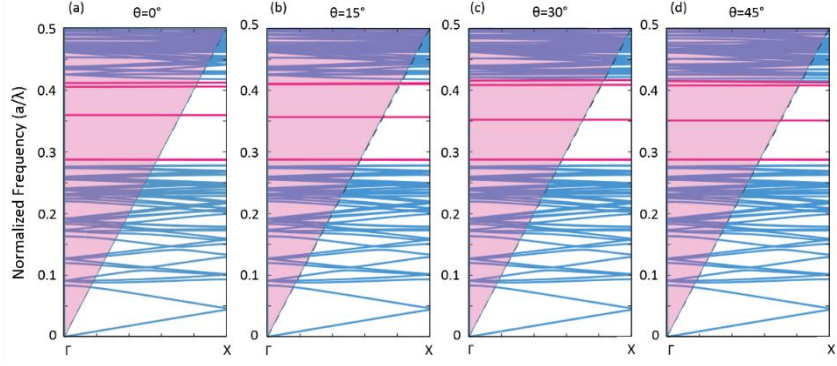

**Figure 2.** Band diagrams of the $(5a \times 5a)$ supercell according to the rotation angle of the small rod within defect region, $\theta_{Rot} = 0°$ (a), $\theta_{Rot} = 15°$ (b), $\theta_{Rot} = 30°$ (c) and $\theta_{Rot} = 45°$ (d).

*2.2 Time domain analysis*

We introduce transmission peaks for each of rotational angle within a broad range to realize the mode shifts in wavelengths, these analyses have repeated for all existing modes for the designed structure. The mode identification has performed in an exact way with parallel analyses of electric field distributions, that we will discuss later. Without any hesitation, a big portion of peak shifting is observed at the second mode ($M_2$) which is used for telecom wavelength, which can be given as $\lambda_2 = [1561\ nm, 1571\ nm, 1589\ nm, 1597\ nm]$ for $\theta_{Rot} = [0°, 15°, 30°, 45°]$ rotational angles, respectively with the maximum deviation is $\lambda_{shift_2} = [1561\ nm,\ 1597\ nm]$ which corresponds to $36\ nm$ shifting between $\theta_{Rot} = [0° - 45°]$ angles, is represented in figure 3(b). A similar trend exists for the third mode ($M_3$) which has a wavelength range as $\lambda_3 = [1381\ nm, 1366\ nm, 1346\ nm, 1355\ nm]$ for $\theta_{Rot} = [0°, 15°, 30°, 45°]$, respectively. A reasonable shift is valid between $\theta_{Rot} = [0° - 30°]$ rotational angle deviation, and maximum peak shift $\lambda_{shift_3} = [1346\ nm,\ 1381 nm]$ with $35\ nm$ shifting value, results are given in figure 3(c). Between the modes that show nearly same attitude to peak shifting with respect to rotational alteration ($M_2\ and\ M_3$), the existence of a common behavior of some wavelengths towards higher rotational angles is also provided an insight for the mode evaluation of wavelengths. However, there is almost no shift for the first mode ($M_1$) which has a range as $\lambda_1 = [1956\ nm, 1956\ nm, 1956\ nm, 1955\ nm]$ whilst a slight shifting exists for the fourth mode ($M_4$) that belongs to the following wavelengths $\lambda_4 = [1362\ nm, 1365\ nm, 1372\ nm, 1375\ nm]$, for $\theta_{Rot} = [0°, 15°, 30°, 45°]$, respectively. The slight change of $13\ nm$ for $M_4$ exists between $\theta_{Rot} = [0° - 45°]$ with the approximate values of $\lambda_{shift_4} = [1362\ nm,\ 1375\ nm]$ for another telecom wavelength which is illustrated in figures 3(a) and 3(d), respectively. By considering the various type of wavelength tuning cavity studies in terms of modifications; position [28,29], thickness [30], material [31] etc., it can be claimed that our study, which uses the rotational modification in the cavity region, at a good position without any drop in quality performance. As expected from the band diagrams, wavelengths are consistent to each other at every used rotational angle for band diagrams and transmission spectrum outcomes. In addition to frequency domain investigations of the cavity, we also focus on time domain analyses as an additional tool requirement for $\boldsymbol{H}$-polarization which is considered as transverse magnetic (TM) mode. In the scope of time domain study, some performance criterions like quality and Purcell factors should be taken into consideration. Here, electric field distributions of the modes for all exists for the designed structure are given with quality and Purcell values regarding $\theta_{Rot} = [0°, 15°, 30°, 45°]$ rotational angles for $\lambda_1 = [1956\ nm, 1956\ nm, 1956\ nm, 1955\ nm]$, $\lambda_2 = [1561\ nm, 1571\ nm, 1589\ nm, 1597\ nm]$, $\lambda_3 = [1381\ nm, 1366\ nm, 1346\ nm, 1355\ nm]$, and $\lambda_4 = [1362\ nm, 1365\ nm, 1372\ nm, 1375\ nm]$ wavelengths in figure 4, respectively. Quality factor is the decaying rate of electric field in a cavity region, it can be identified as a criterion to determine the performance of the light confinement within the cavity. If we evaluate in the way of quality, we can describe our designed cavity as a high quality microcavity, according to the results that can be shown in figure 4(a). We obtained high quality values for each of configurations by performing finite difference time domain method (FDTD), but the most satisfying results are obtained for $M_2$ with high values of $Q = [2.296x10^8, 1.997x10^8, 1.405x10^8, 1.043x10^8]$ by showing a decaying trend with regard to increasing rotational angle that can we relate the angles as $\theta_{Rot} = [0°, 15°, 30°, 45°]$, respectively. As can be seen from the obtained values the highest quality factor value in 2D is $Q = 2.296x10^8$, and it exists at $\theta_{Rot} = [0°]$. Relatively small values are obtained for other modes $M_1$ and $M_3$ in consequence of bigger mode areas by comparison to $M_2$, an intuitive thought can be formed from figure 4(b) [13,32]. The obtained high 2D quality factor values are comparable with the other microcavity studies [13,16,33–35]. Purcell factor is another figure of merit, which measures the radiation enhancement rate of the [36] cavity region. This property is also calculated by using LUMERICAL FDTD method for all configurations by one by, the outcomes can be seen from figure 4(a) [37]. The results show that magnitudes of the Purcell factors are correlated with quality factors on the basis of best fit wavelength. However, the rotation angle dependency of the Purcell factor does not give results in parallel with those of the quality. Higher Purcell value means that much better field enhancement is provided, by using this information, $F_{p2} = 331.412$ value at the rotational angle of $\theta_{Rot} = [15°]$ is better in comparison to other values obtained. The fact that there is not much difference between the Purcell and quality factor values for the varying rotational angles indicates that the area of the light confinement is close to each other for each angle. This manner is valid for each mode in itself. The Purcell factors at $M_2$ for each rotational angle $\theta_{Rot} = [0°, 15°, 30°, 45°]$ can be ordered as follows, respectively; $F_p = [281.165, 331.412, 325.910, 310.605]$. Based on the mode profiles, we investigate E-field distributions and obtained the results that is illustrated in the figure 4. According to the outcomes, it can be referred that the light strongly confines on some specific points which are around the defect region, the claimed points correspond to the dielectric rods substantially. The localization on the dielectric rods shows identical e-field distributions, which are called as modes and exist at separate wavelengths. We come across four types of modes for our designed structure, these modes are supported by our cavity structure at some exact wavelengths and described above as $M_1$, $M_2$, $M_3$ and, $M_4$. All the supported modes are observed at the band diagrams and transmission peak shifts also, this accord between our analyses gives a clue about the way that we are on is true. The most reasonable results are obtained for the $M_2$, this mode takes part in a location which is far away from the edge of the band gap for all applied defect region modifications. This type of positioning is an inducement for showing the ideal attitude of mentioned wavelength for all cases. The other operating wavelengths that belongs to $M_3$ and, $M_4$ are observed around the positions of band gap edge. The destructive effects of the edge conditions are violently sensed for the cases of the nearest positioning exists on the $\theta_{Rot} = [30°, 45°]$ rotational angles for mentioned modes. The drop of cavity effect and the lack of the localization attitude of the modes are an expected consequence of getting closer to the band edge.

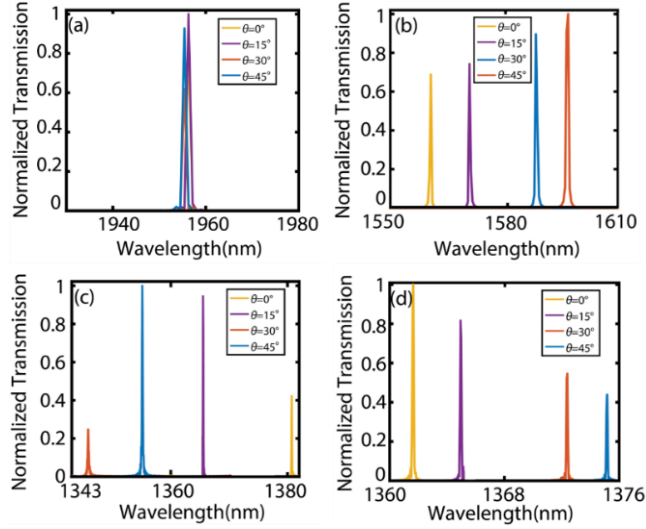

**Figure 3.** Normalized transmission peak shift modes $M_1$, (b) $M_2$, (c) $M_3$, (d) $M_4$ according to rotation angles.

Investigations show that the most reasonable results are obtained for the $M_2$ which is supported by the wavelengths as $\lambda_2 = [1561\ nm, 1571\ nm, 1589\ nm, 1597\ nm]$ for $\theta_{Rot} = [0°, 15°, 30°, 45°]$ rotational angles. Therefore, it is decided that to focus for deeper investigations on this wavelength and three-dimensional (3D) analyses are carried out by considering finite thicknesses of the structure long the third dimension specifically out-of-plane direction.

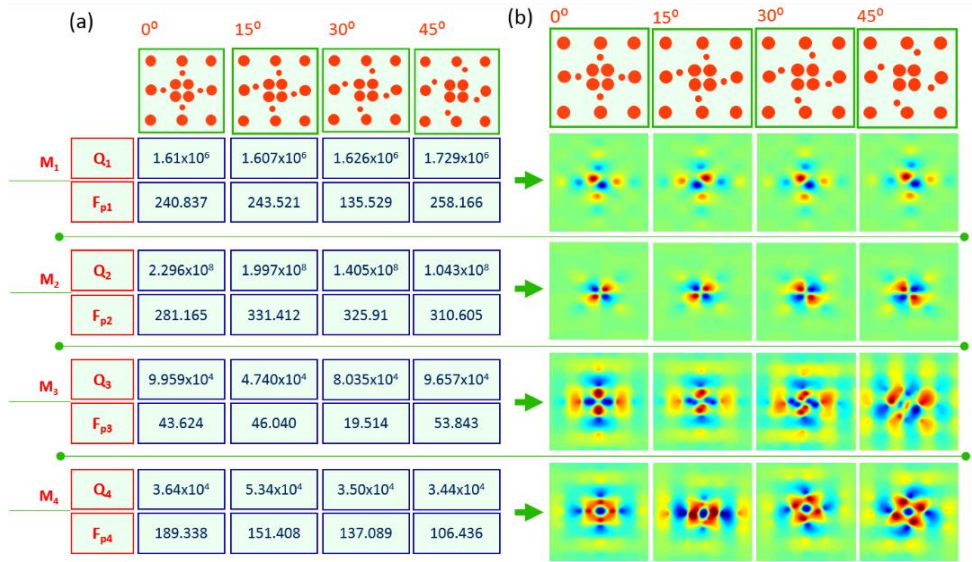

**Figure 4.** (a) Presentation of Quality factor $Q = [Q_1, Q_2, Q_3, Q_4]$ and Purcell factor $F_p = [F_{p1}, F_{p2}, F_{p3}, F_{p4}]$ values according to rotational angles, subscripts indicates the first ($M_1$), second ($M_2$), third ($M_3$) and fourth modes ($M_4$) of the designed structures, respectively. (b) E-field distributions of each modes with respect to rotational angles $\theta_{Rot} = [0°, 15°, 30°, 45°]$.

Before presenting the 3D analyses of cavity structure, it should be mentioned briefly the mode volume, which is related to Purcell and quality factors. To emphasize the relationship between these figure of merits, it will be beneficial to present the correlation equation which is called the Purcell factor [36].

$$F_p = \frac{3}{4\pi^2}\left(\frac{\lambda_c}{n_c}\right)^3\left(\frac{Q}{V_{eff}}\right), \qquad (1)$$

where, $\lambda_c$ is the free space wavelength, $n_c$ is the corresponding refractive index of the using material in the cavity region at the operation wavelength, $Q$ is the quality factor, and $V_{eff}$ is the effective mode volume where the light localized. According to equation (1), it is clear that Purcell factor has a linear relation with quality factor while it is inversely proportional to mode volume. As an important expectation of the cavity performance, mode volume should be as small as possible while the quality factor should have high values to fulfill the requirement of high Purcell factor, which is the enhancement parameter of the cavity region. Here, we present the performance results of the designed cavity for one-layer 3D simulations by using FDTD simulations for each rotational angle at the mode $M_2$, which gives the best results for prior analyses in table 1. It can be easily seen that, the light confines in a small volume as $V\ (\lambda/n)^3 = [2.9, 2.5, 2.2, 2.3]$ and has respectively small qualities factors $Q = [1263.5, 1248.5, 1216.0, 1260.5]$ for the rotational angles $\theta_{Rot} = [0°, 15°, 30°, 45°]$, respectively.

These results can be commented as small mode volumes by quoting references to cavity studies in the literature [6,13,38–40]. However, the quality and Purcell factor values are not as higher as those of the 2D.

**Table 1.** Quality $Q$, mode volume $V$ $(\lambda/n)^3$ and Purcell factors $F_p$ of silicon for mode $M_2$.

|  | $Q$ | $V$ $(\lambda/n)^3$ | $F_p$ |
|---|---|---|---|
| $\theta_{Rot_0}$ (°) | 1263.5 | 2.9 | 33.3 |
| $\theta_{Rot_{15}}$ (°) | 1248.5 | 2.5 | 37.6 |
| $\theta_{Rot_{30}}$ (°) | 1216.0 | 2.2 | 41.4 |
| $\theta_{Rot_{45}}$ (°) | 1260.5 | 2.3 | 41.9 |

The reason for the dramatic lower quality factor results of the 3D compared to 2D is the leakage of the light into the air from the top and bottom interfaces of the structure, as it is mentioned in the introduction part. This leakage is caused by the inability of the light satisfying TIR condition between the layers along the z-axis, therefore it leaks instead of getting trapped inside the cavity region. To define a structure as 2D, it is required to compare the operating wavelength and the structural length on the basis of the investigated axis. A structure can be assumed 2D as long as the mentioned dimensional length is larger enough than the operating wavelength by approaching an infinite value. With reference to dimensional approximation, it can be alleged that a 2D structure can be obtained by stacking 3Ds. By taking advantage of this clue, we also numerically performed analysis of 3D structures to mimic 2D counterpart. The ideal case in terms of wavelength and rotational angle, which the best quality factor value is obtained has preferred for application the stacking method of the structures in the z-axis. According to results, it seems that we are following an appropriate way to reach the 2D structure because an exponentially increasing trend of the quality factor appears whilst the number of layers increases. The quality factors of $\lambda_2 = 1561\ nm$ are shown in the figure 5 for $\theta_{Rot} = [0°]$ rotational angle, conforming to the number of layers. The quality factors are $Q = [1263, 4702, 10326]$, with respect to increasing number of layers. When the second order fit is applied to the quality factor results, it is recognized that values are reached to $2.296 \times 10^8$ 2D quality factor by providing 3D structure when there are approximately 450 layers ($2250\ \mu m$). In terms of dimensional 2D approximation, 450 stack layer of 3D structure corresponds to 1441 times wavelength. This verify converges to the nearly infinite z-axis with respect to operating wavelength and verifies 2D approach by stacking 3D structure.

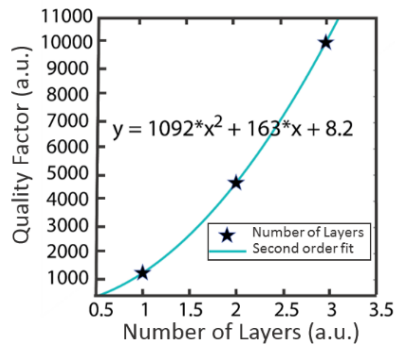

**Figure 5.** Quality factor trend of 3D structure with respect to increasing number of layers.

*2.2 Additional Frequency Domain Analysis*

To be able to compare the numerical mode analyses with real world conditions, experimental analysis has been needed. When taking into consideration the experimental setup that includes alumina rods, new numerical analysis implementation obligation has been raised to alumina material $\varepsilon_r = 9.61$ for *H*-polarization. Thus, a new opportunity has emerged in terms of the evaluation of two different dielectric material ($\varepsilon_r = 12$ and $\varepsilon_r = 9.61$) behavior for the same cavity structure. Unlike the silicon dielectric material, the designed cavity structure with alumina supports only two modes. The disappearance of the rest of the modes, which has emerged near bandgap, can be explained as a shifting of these modes towards to the band region. According to the given numerical transmission results of the structure consisting of the $\varepsilon_r = 9.61$ rods are given in figure 6(a), transmission peaks appear at $\lambda = [1770\ nm_{\lambda_1}, 1418\ nm_{\lambda_2}]$ wavelengths and these peaks can be represented by $M_1$ and $M_2$, respectively. It can be seen that two modes exist with approximately $200\ nm$ shift to the lower wavelengths with respect to silicon at $\theta_{Rot} = [0°]$ as a simple configuration. The difference of the mode existing wavelengths between silicon and alumina, can be associated with the refractive index ratio. In terms of the quality factors of mentioned wavelengths, the quality factors indicate a similar attitude with silicon material and the results show that $Q = [6.742x10^5{}_{Q_1}, 1.607x10^7{}_{Q_2}]$ for $\lambda_1 = 1770\ nm$ and $\lambda_2 = 1418\ nm$, respectively. As it performed in silicon case, the ideal mode determination has been done based on high quality and wavelength tunability, for two wavelengths in alumina. As a result of this investigation, the ideal mode is determined as $M_2$. After that, wavelength tunability has been analyzed for the ideal mode at $\theta_{Rot} = [0°, 15°, 30°, 45°]$ rotational angles, see figure 6(b).

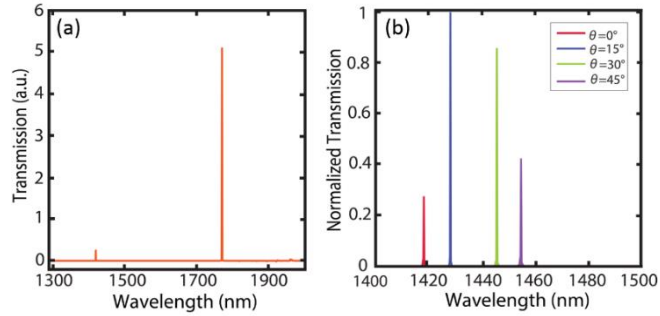

**Figure 6.** (a) Numerical transmission graph for existing modes at $\theta_{Rot} = [0°]$, (b) transmission graphs at each rotational angles $\theta_{Rot} = [0°, 15°, 30°, 45°]$ for $M_2$ mode.

According to the analyses, results for ideal mode show that there is a significant wavelength tune is possible by means of designed cavity structure, $\lambda_2 = [1418\ nm, 1428\ nm, 1446\ nm, 1455\ nm]$, for $\theta_{Rot} = [0°, 15°, 30°, 45°]$, respectively. The wavelength change of $37\ nm$ for $M_2$ exists between $\theta_{Rot} = [0° - 45°]$ with the approximate values of $\lambda_{shift_2} = [1418\ nm, 1455\ nm]$. Furthermore, quality analyses of ideal mode at specified rotational angles are also conducted, and results in $Q = [1.607x10^7, 1.562x10^7, 1.419x10^7, 1.209x10^7]$ for $\theta_{Rot} = [0°, 15°, 30°, 45°]$, respectively. Besides 2D simulations, 3D analyses are performed in terms of quality and Purcell factors to compare the results with that of experimental ones. The results show that the cavity structure confines the light for the ideal mode $M_2$ in a similar way as the silicon results, see table 2. Quality and Purcell factors can be represented as, $Q = [1598.9, 1452.2, 1374.9, 1471.8]$ and $V(\lambda/n)^3 = [3.0, 2.8, 2.6, 2.9]$ at $\theta_{Rot} = [0°, 15°, 30°, 45°]$ rotational angles, respectively.

**Table 2.** Quality $Q$, mode volume $V(\lambda/n)^3$ and Purcell factors $F_p$ of alumina for mode $M_2$.

|  |  | Q | $V(\lambda/n)^3$ | $F_p$ |
|---|---|---|---|---|
| $\theta_{Rot_0}$ | (°) | 1598.9 | 3.0 | 40.9 |
| $\theta_{Rot_{15}}$ | (°) | 1452.2 | 2.8 | 39.0 |
| $\theta_{Rot_{30}}$ | (°) | 1374.9 | 2.6 | 39.8 |
| $\theta_{Rot_{45}}$ | (°) | 1471.8 | 2.9 | 38.2 |

As can be seen from table 2, Purcell and quality factors and also mode volumes are the approximate values that of silicon, additionally the quality factor has taken their highest value at $\theta_{Rot} = [0°]$. All the analyses indicate that alumina and silicon show a similar performance attitude except for wavelength operation range for our designed structure.

## 3. Experimental Study

After the numerical investigation is conducted, microwave experiments are performed to verify the overlapping between the real enviroment and our design. In the prepared setup that is for performing transmission measurement, Agilent E5071C type network analyzer, standard horn antennas as receiver and transmitter are used. While transmitter and receiver antennas are positioned outside of the center axis of the $15x15$ sized PhC structure to adequate distances to be able to diagnose the operation wavelengths which cavity effect has emerged, the measurement position of the transmitter can be defined as a near-field range, see figure 7(a). Cylindrical alumina rods that have $\varepsilon_r = 9.61$ dielectric constant, $R_1 = 3.17\ mm$ $R_2 = 6.35\ mm$ diameters, and $h = 15.3\ cm$ height are utilized in the PhC structure, these rods construct a square structure with dimensions, $w = 23.8\ cm$ widths and $l = 23.8\ cm$ lengths and according to the given dimensions the lattice constant calculated as $a = 15.85\ mm$.

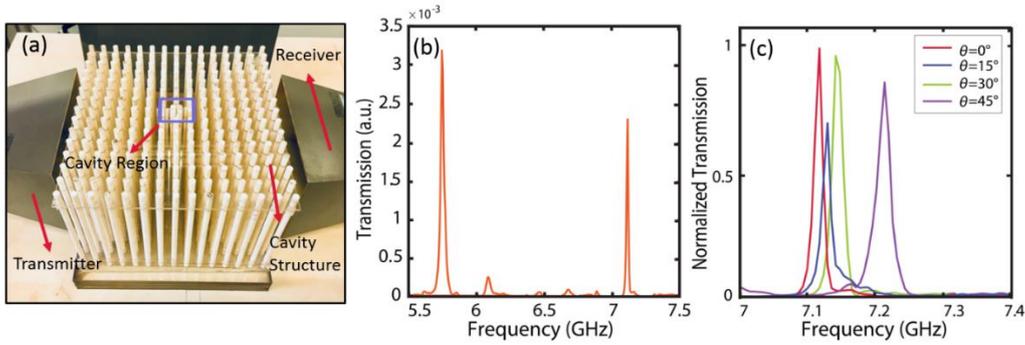

**Figure 7.** (a)Schematic representation of microwave setup with the blue square demonstration of cavity region, (b) experimental transmission graphs for each existing cavity modes according to $\theta_{Rot} = [0°]$ rotational angle, (c) central peak shifts of ideal mode $M_2$ with respect to $\theta_{Rot} = [0°, 15°, 30°, 45°]$ rotational angles.

Experimental operation range is calculated based on the numerical cavity mode wavelengths. conforming to wavelength-frequency transition, predicted operation frequency interval is $6\ GHz - 7.5\ GHz$. As it is mentioned above experimental measurements are given at $\theta_{Rot} = [0°]$ due to aiming the main goal in a simple configuration without any performance metrics like wavelength tuning and, quality factor. In this manner, we could compare experimental results with 3D stacking layer approximation also. Transmission measurement results in microwave regime is given in figure 7(b), experimental center frequency results $f = [5.746\ GHz_{f_1}, 7.117\ GHz_{f_2}]$ are correlated with that of numerical ones at $\theta_{Rot} = [0°]$. It should be also mentioned, there is a slight frequency deviation $f_{dev} = [0.25\ GHz - 0.37\ GHz]$ interval which is caused by the environmental perturbation and array issues of the structure, also possible irregularities of the rods can be added as a reason.

To find out a way from general to specific, we focused on the most utilizable wavelength in terms of the wavelength tunability and quality factor. There is a need to investigate the experimental reaction of the modes to angular modifications, for comparableness. In addition to previous angular modification response of the cavity modes numerically, experimental mode analyses are performed applying transmission measurement on the PhC structure. Ideal mode $M_2$ shows $f = [7.117\ GHz_{f_{0°}}, 7.129\ GHz_{f_{15°}}, 7.141\ GHz_{f_{30°}}, 7.213\ GHz_{f_{45°}}]$ transmission peaks responding to $\theta_{rot} = [0°, 15°, 30°, 45°]$ rotational angles, see figure 7(c). Similar to the numerical results, maximum shifting is observed between $\theta_{rot} = [0°, 45°]$ interval with $0.0960\ GHz$ shifting value for $f_{shift} = [7.117\ GHz, 7.213\ GHZ]$ frequencies. After the ideal mode gives satisfying wavelength tunability results the investigation of quality factor emerges as a requirement for a efficient cavity effect. On the contrary of numerical qulity factor investgtion, low quality analysis is used obtaining experimental results, low quality analysis formula is represented in equation (2) [41].

$$Q = \frac{f_{center}}{f_2 - f_1}. \qquad (2)$$

Equation (2) includes $f_{center}$ and $f_2 - f_1$ as representing central value of operating frequency and bandwidth of the curve, respectively. All the existing modes are analyzed to be sure that the ideal mode is also valid in terms of quality factor like wavelength tunability. According to the results, $Q = \left[212.02_{Q_1}, 492.2_{Q_2}\right]$, the ideal mode is compatible with numerical ones. With high wavelength tunability and quality factor, $M_2$ is approved as the most efficient one, this mode shows $Q = \left[492.2_{Q_{0°}}, 502.7_{Q_{15°}}, 378.6_{Q_{30°}}, 330.0_{Q_{45°}}\right]$ quality factors for $\theta_{rot} = [0°, 15°, 30°, 45°]$ angular modifications, respectively. Additionally to the comparison between 2D and 3D in terms of quality factors, a similar analogy can apply for 3D and experimental quality values, for achieving further information about the behavior of PhC structure. According to the outcomes, there is an appreciable consistency for each rotation angle for 3D and experimental results. Even though the mentioned results show a decrease with respect to simulation 3D results, this discrepancy can be described as environmental effects which are caused by the structural imperfections of the cavity or active regions. These can be sorted as the unequal sequence, tilt, or imperfections of rods [14].

## 4. Conclusion

This study shows a roadmap to achieve the most ideal case for a cavity mode by investigating the performance in terms of various criteria and also includes a comparison between two different dielectric materials in the manner of cavity ability. Silicon takes part in the center of the study, and the investigations are conducted with frequency and time domain analyses. For silicon based structure frequency domain analysis exhibits that obtained four resonance modes shiftings have been had respectively better results for specified rotational angles $\theta_{rot} = [0°, 15°, 30°, 45°]$ according to other studies. By using time domain analysis, such a high, ~$10^8$, quality is achieved for 2D. In terms of 3D analysis, an ideal mode has been chosen in terms of providing a maximum wavelength shift 36nm, and the highest quality factor $2.296x10^8$ at $\theta_{rot} = 0°$, among to the other modes. According to 3D results, quality factor decrease has been existed to 1263.5, as expected. To perform the experimental investigation these analyses have been repeated for alumina and quite similar outcomes have been acquired such as $1.607x10^7$ quality factor and 37 nm wavelength shift. Additionally, the 3D analyses have been conducted for quality performance, Purcell factor and mode volume investigation to be able to do a reasonable comparison with silicon. Then, the experimental stage was carried out for two cavity modes, ideal mode selection was compatible that of numerical in terms of the quality factor and wavelength shift 492.2 and $0.0960\ GHz$, respectively. All the results show that our respectively compact and simple designed microcavity structure provides high tunability and control on several modes which operate on telecom wavelengths mostly.

**Acknowledgements**

The authors M.G. and H.K. greatfully acknowledge the financial support of the Scientific and Technological Research Council of Turkey (TUBITAK) with Project No. 115R036. H.K. also acknowledges the partial support of the Turkish Academy of Sciences.